\documentclass[
    ,final            
]
  {aipproc}
  \usepackage{amssymb,amsmath,tabularx,float}

\layoutstyle{8x11double}

\def\beq{\begin{equation}}
\def\eeq{\end{equation}}
\def\beqr{\begin{eqnarray}}
\def\eeqr{\end{eqnarray}}
\def\bdpm{\begin{displaymath}}
\def\edpm{\end{displaymath}}
\def\npb#1#2#3 {Nucl. Phys. B {\bf#1}, #2 (#3)}
\def\plb#1#2#3 {Phys. Lett. B {\bf#1}, #2 (#3)}
\def\prd#1#2#3 {Phys. Rev. D {\bf#1}, #2 (#3)}
\def\jhep#1#2#3 {J. High Energy Phys. {\bf#1}, #2 (#3)}
\def\jpg#1#2#3 {J. Phys. G {\bf#1}, #2 (#3)}
\def\epj#1#2#3 {Eur. Phys. J. C {\bf#1}, #2 (#3)}
\def\arnps#1#2#3 {Ann. Rev. Nucl. Part. Sci. {\bf#1}, #2 (#3)}
\def\ibid#1#2#3 {{\it ibid.} {\bf#1}, #2 (#3)}
\def\none#1#2#3 {{\bf#1}, #2 (#3)}

\begin{document}

\title{Fermions and gauge bosons in $SU(4)_L \times U(1)_X$ models with little Higgs}


\author{Soo-hyeon Nam}{
  address={Department of Physics, National Cheng-Kung
University, Tainan 701, Taiwan \\
National Center for Theoretical Sciences, Hsinchu 300, Taiwan}
}

\begin{abstract}
We discuss the aspects of the little Higgs model with the $SU(4)_L \times U(1)_X$ electroweak gauge group as an alternative solution to the naturalness and fine-tuning issues.  We introduce anomaly-free fermion spectra, and present their interactions with the physical gauge bosons. We also discuss some phenomenological implications of these fermions and the extra gauge bosons based on recent experimental results.
\end{abstract}

\maketitle


The little Higgs model (LHM) based on an $SU(4)_L\times U(1)_X$ gauge group was recently proposed by Kaplan and Schmaltz as an alternative solution to the naturalness and fine-tuning issues \cite{Kaplan03}. The LHM adopts the early idea that Higgs can be considered as a Nambu Goldstone boson from global symmetry breaking at some higher scale $\Lambda \sim 4\pi f$ \cite{Dimopoulos82} and acquires a mass radiatively through symmetry breaking at the electroweak scale $v$ by collective breaking \cite{Arkani01}.  The LHM with the $SU(4)_L\times U(1)_X$ symmetry appears fundamentally different from other types of LHMs due to the multiple breaking of global symmetry by separate scalar fields \cite{Schmaltz05}.  In this model, the bound on the new symmetry breaking scale $f$ of the  gauge group was obtained earlier from the tree-level electroweak constraints  in Ref. \cite{Csaki03}.  This model has vector-like heavy quarks of charge 2/3 for each generation with a simple family universal embedding, but this choice leaves nonvanishing gauge anomalies which require additional fermion multiplets at the scale $\Lambda$.  Instead of adding phenomenologically ambiguous extra fermions, one can construct a fermion sector that is already anomaly-free at the scale $f$, which can be done by embedding the first two generations of quarks into $\bar{4}$ representations of $SU(4)_L$ while the third generation of quarks and all three generations of leptons are into $4$s of $SU(4)_L$ \cite{SU4,Otto03}. With this anomaly-free choice of the fermions, we reexamine the electroweak precision constraints on this model.

The scalar sector in this model is based on the non-linear sigma model describing $[SU(4)/SU(3)]^4$ global symmetry breaking with the diagonal $SU(4)$ subgroup gauged and four non-linear sigma model field $\Phi_{ij}$ where $i,j = 1,2$.
The standard $SU(2)_L\times U(1)_Y$ gauge group can be embedded into the theory with an additional $U(1)_X$ group.  The $SU(4)$ breaking is not aligned and only the gauged $SU(2)$ is linearly realized in this model where the scalar fields $\Phi_{ij}$ can be parametrized as \footnote{Our notation on the VEVs is different from that of Kaplan and Schmaltz. For instance, $f_2$ in this letter equals to $f_{34}/\sqrt{2}$ shown in Ref. \cite{Kaplan03}.  But the scale parameter $f$ is the same eventually.}
\beq
\Phi_{11}=e^{+i {\cal H}_u \frac{f_{12}}{f_{11}}}
        \left( \begin{array}{l} 0  \\ 0 \\ f_{11} \\0 \end{array} \right) ,\
\Phi_{12}=e^{- i {\cal H}_u \frac{f_{11}}{f_{12}}}
        \left( \begin{array}{l} 0  \\ 0 \\ f_{12} \\0 \end{array} \right) ,
        \nonumber
\eeq
\beq
\Phi_{21}=e^{+i {\cal H}_d \frac{f_{22}}{f_{21}}}
        \left( \begin{array}{l} 0  \\ 0 \\ 0 \\f_{21} \end{array} \right) ,\
\Phi_{22}=e^{- i {\cal H}_d \frac{f_{21}}{f_{22}}}
        \left( \begin{array}{l} 0  \\ 0 \\ 0 \\f_{22} \end{array} \right) ,
\eeq
where
\beq
  {\cal H}_u =
    \left( \begin{array}{ccc}
           \begin{array}{cc} 0 & 0 \\ 0 & 0 \end{array}
             & h_1 & \begin{array}{c} 0 \\ 0 \end{array}   \\
            h_1^\dagger & 0 &  0 \\
           \begin{array}{cc}  0 &  0 \end{array} & 0 & 0 \\
           \end{array} \right)\Big{/}2f_{1} , \nonumber
\eeq
\beq
  {\cal H}_d =
    \left( \begin{array}{ccc}
      \begin{array}{cc} 0 & 0 \\ 0 & 0 \end{array}
      &  \begin{array}{c}  0 \\  0 \end{array}  &  h_2 \\
        \begin{array}{cc}  0 &  0 \end{array}  & 0 &  0 \\
      h_2^\dagger &  0 & 0 \\
    \end{array} \right)\Big{/}2f_{2} ,
\eeq
and $f^2_{i} = \frac{1}{2}\sum_{j=1,2} f^2_{ij}$.  Here we only show the two complex doublets $h_{1,2}$ and discard the other singlets whose contributions to masses are negligible.

 The covariant derivative of the scalar and fermion quadruplets is given by
\beq
D_\mu=\partial_\mu + ig T_L^\alpha A_\mu^\alpha + ig_X X A^X_\mu ,
\eeq
where $X$ is the $U(1)_X$ charge, $A_\mu^\alpha, g$ and $A^X_\mu, g_X$ are the gauge bosons and couplings of the $SU(4)_L$ and $U(1)_X$ gauge groups, respectively, and $T_L^\alpha = \lambda^\alpha/2$ with $\lambda^\alpha$ the Gell-Mann matrices for $SU(4)_L$ normalized as Tr$(\lambda^\alpha\lambda^\beta) = 2\delta^{\alpha\beta}$.
The electric charge generator is a linear combination of the $U(1)_X$ generator and the three diagonal generators of the $SU(4)_L$ group:
\beq
\mathnormal{Q}=a_1T^3_{L}+\frac{a_2}{\sqrt{3}}T^8_{L}+ \frac{a_3}{\sqrt{6}}T^{15}_L+ XI_4 ,
\eeq
with
\beq
T_L^3 = \frac{1}{2}\textrm{diag}(1,-1,0,0),\ T_L^8 = \frac{1}{2\sqrt{3}}\textrm{diag}(1,1,-2,0), \nonumber
\eeq
\beq
T_L^{15} = \frac{1}{2\sqrt{6}}\textrm{diag}(1,1,1,-3), \ I_4 = \textrm{diag}(1,1,1,1).
\eeq
The free parameters $a_1$, $a_2$ and $a_3$ fix the electroweak charges of the scalar and fermion representations.  In particular, $a_1=1$ gives the correct embedding of the standard model (SM) isospin $SU(2)_L$ doublets ($Q = T_L^3 + Y$).
In this model, the standard $SU(2)_L\times U(1)_Y$ ($t,b$) doublet is embedded into a $SU(4)_L\times U(1)_X$ quadruplet $\psi_L$ as follows
\beq
\psi_L = (t, b, T, B)_L^T.
\eeq
Here we choose $T(B)$ to have the same electric charge as $t(b)$, and
the duplicated extra heavy fermions ($T,B$) remove the quadratic divergences due to their SM fermion partners ($t,b$). The full anomaly-free fermion spectrum for three families can be found in Ref. \cite{Otto03}. In this case, the parameters $a_2$ and $a_3$ are uniquely determined as $a_2=-1$ and $a_3=2$, and the $X$ charges of the scalar and quark multiplets are given as
\beq
X(\Phi_{1j}) = -\frac{1}{2},\, X(\Phi_{2j}) = \frac{1}{2},\, X(\psi_L) = \frac{1}{6},\, X(\psi_R) = \mathnormal{Q},
\eeq
and the VEVs of the two doublet Higgs fields $h_{1,2}$ are of the following form
\beq
\langle h_1\rangle = {v_u \choose 0} , \quad \langle h_2 \rangle = {0 \choose v_d}.
\eeq
Note that our quantum number assignments for the scalars and fermions as well as the form of the Higgs fields differ from Ref. \cite{Kaplan03} due to the different choice of the fermion multiplets.

\begin{table}[hbt!]
\begin{tabular}{lllll}
\hline
  $\psi$ & $g_V$ & $g_A$ & $g_V^\prime$ & $g_A^\prime$ \\
  \hline
  $t$ & $\frac{1}{2}-\frac{4}{3}s_W^2 +\frac{5t}{6} s_ws_\theta$     &   $\frac{1}{2} -\frac{t}{2}s_Ws_\theta $ &  $\left(\frac{1}{2}-\frac{4}{3}s_W^2\right)s_\theta-\frac{5t}{6}s_W$ & $\frac{1}{2}s_\theta + \frac{t}{2}s_W $ \\
  $b$ &  $-\frac{1}{2}+\frac{2}{3}s_W^2 -\frac{t}{6} s_ws_\theta$     &   $-\frac{1}{2} +\frac{t}{2}s_Ws_\theta $   &  $\left(-\frac{1}{2}+\frac{2}{3}s_W^2\right)s_\theta +\frac{t}{6}s_W$ & $-\frac{1}{2}s_\theta - \frac{t}{2}s_W $ \\
  $\nu$ &  $\frac{1}{2} -\frac{t}{2} s_ws_\theta$    &   $\frac{1}{2} -\frac{t}{2}s_Ws_\theta $  &  $\frac{1}{2}s_\theta +\frac{t}{2}s_W$ &  $\frac{1}{2}s_\theta + \frac{t}{2}s_W $ \\
  $e$ &  $-\frac{1}{2} +2s_W^2 -\frac{3t}{2} s_ws_\theta$  &   $-\frac{1}{2} +\frac{t}{2}s_Ws_\theta $   & $\left(-\frac{1}{2} + 2s_W^2\right)s_\theta +\frac{3t}{2}s_W$  &  $-\frac{1}{2}s_\theta - \frac{t}{2}s_W $ \\
\hline
\end{tabular}
\caption{$Z$ and $Z'$ couplings to the SM fermions}
\label{tab:couplings}
\end{table}

Since each of $a_i$ is chosen, the gauge boson structure of the electroweak sector is fixed, and the 15 gauge fields $A^\alpha_\mu$ associated with $SU(4)_L$ can be written as
\beq
T_L^\alpha A^\alpha_\mu=\frac{1}{\sqrt{2}}\left(
\begin{array}{cccc}Z^{0}_{1\mu} & W^{+}_\mu & Y^{0}_\mu & X^{\prime +}_\mu\\
W^{-}_\mu & Z^{0}_{2\mu} &  X^{-}_{\mu} &  Y^{\prime 0}_\mu\\
\bar{Y}^{0}_\mu & X^{+}_{\mu} & Z^{0}_{3\mu} & W^{\prime +}_\mu\\
X^{\prime -}_\mu & \bar{Y}^{\prime 0}_\mu & W^{\prime -}_\mu & Z^{0}_{4\mu} \end{array}\right),
\eeq
where
$Z_{1\mu}^0=A^3_\mu/\sqrt{2}+A^8_\mu/\sqrt{6}+A^{15}_\mu/\sqrt{12}$,
$Z_{2\mu}^0=-A^3_\mu/\sqrt{2}+A^8_\mu/\sqrt{6}+A^{15}_\mu/\sqrt{12}$,
$Z_{3\mu}^0=-2A^8_\mu/\sqrt{6}+A^{15}_\mu/\sqrt{12}$, and
$Z_{4\mu}^0=-3 A^{15}_\mu/\sqrt{12}$.
After spontaneous symmetry braking, the charged gauge bosons have the following mass terms:
\bdpm
M^2_{W} = \frac{1}{4}g^2v^2, \quad M^2_{W^{\prime}} = \frac{1}{4}g^2\left(4f^2 - v^2 \right),
\edpm
\bdpm
M^2_{X} = \frac{1}{4}g^2\left(4f_1^2 - v_1^2 + v_2^2 \right),\quad M^2_{Y} = g^2f_1^2,
\edpm
\beq
M^2_{X^\prime} = \frac{1}{4}g^2\left(4f_2^2 + v_1^2 - v_2^2 \right),\quad M^2_{Y^\prime} = g^2f_2^2,
\label{Wmass}
\eeq
where
\bdpm
v_1^2 = v_u^2 - \dfrac{v_u^4}{12f_1^2}\left(\dfrac{f_{12}^2}{f_{11}^2}+ \dfrac{f_{11}^2}{f_{12}^2}-1\right),
\edpm
\bdpm
v_2^2 = v_d^2 - \dfrac{v_d^4}{12f_2^2}\left(\dfrac{f_{22}^2}{f_{21}^2}+ \dfrac{f_{21}^2}{f_{22}^2}-1\right),
\edpm
\beq
f^2 = f_1^2+f_2^2, \quad v^2 = v_1^2+v_2^2 ,
\eeq
and for simplicity we use the approximation $f^2 \gg f_1^2-f_2^2,\ v^2 \gg v_1^2-v_2^2$. The three neutral gauge bosons $A^3$, $A^8$ and $A^{15}$ mixing with the $U(1)_X$ gauge boson $A^x$ are associated with a $4\times 4$ nondiagonal mass matrix. After the mass matrix is diagonalized, a zero eigenvalue corresponds to the photon $A$, and the three physical neutral gauge bosons $Z$, $Z'$ and $Z''$ have the following mass terms:
\bdpm
M^2_Z = \frac{g^2v^2}{4c_W^2}\left(1 - \frac{t_W^4}{4}\frac{v^2}{f^2}\right),
\edpm
\beq
M^2_{Z'} = g^2(1+t^2)f^2 - M^2_{Z}, \quad
M^2_{Z''} = \frac{1}{2}g^2f^2 ,
\label{Zmass}
\eeq
where $t \equiv g_X/g$, $c_W \equiv \cos{\theta_W} = \sqrt{(1+t^2)/(1+2t^2)}$, and $\theta_W$ is the Weinberg mixing angle.
Note here that the cancellation of the quadratic divergences in the gauge boson sector is made between the primed and non-primed gauge bosons with $v_i$ in their masses.

For the charged current, the fermion-gauge boson interaction terms (here we only consider the SM fermions in one family) are then given by
{\setlength\arraycolsep{2pt} \beqr
\mathcal{L}_{CC} &=& -\frac{g}{\sqrt{2}}\left[\bar{t}\gamma^\mu(1-\gamma_5)b + \bar{\nu}\gamma^\mu(1-\gamma_5)e\right]W^+_\mu \cr
&&+ (\textrm{terms with}\ X^{(\prime)}, Y^{(\prime)}, \textrm{and}\ W^\prime) + \textrm{h.c}.
\eeqr }
Also, the neutral current is given by
{\setlength\arraycolsep{2pt} \beqr
\mathcal{L}_{NC} &=& - e \mathnormal{Q}\left(\bar{\psi}\gamma^\mu \psi\right)A_\mu +\frac{g}{4\sqrt{2}}\left[\bar{\psi}\gamma^\mu\left(1-\gamma_5\right) \psi\right]Z_\mu^{\prime\prime} \cr
&& - \frac{g}{2c_W}\left[\bar{\psi}\gamma^\mu\left(g_V - g_A\gamma_5\right) \psi\right]Z_\mu \cr
&& - \frac{g}{2c_W}\left[\bar{\psi}\gamma^\mu\left(g^{\prime}_V - g^{\prime}_A\gamma_5\right) \psi\right]Z^{\prime}_\mu,
\eeqr }
where the values of $g_V^{(\prime)}$ and $g_A^{(\prime)}$ are listed in Table \ref{tab:couplings}. One can see from the table that the couplings contain additional new physics (NP) contributions proportional to the mixing angle $\theta$ between $Z$ and $Z'$ where $s_\theta \equiv \sin\theta =   t_W^2\sqrt{1-t_W^2}v^2/(2c_Wf^2)$.

We now calculate the electroweak constraints on this model from several physical observables. From Eq. (\ref{Zmass}), we find the custodial $SU(2)_L$ symmetry violating shift in the $Z$ mass,
\beq
\delta\rho = \alpha T \simeq \dfrac{g_X^4}{4(g^2+g_X^2)^2}\dfrac{v^2}{f^2},
\eeq
where $T$ is the Peskin-Takeuchi parameter which has Higgs mass ($M_H$) dependance.
At 95\% CL for $M_H = 117$ GeV (300 GeV), $T\leq 0.06$ (0.14) has been obtained from various experimental inputs \cite{PDB06}.  This result gives a bound of $f\geq 1.7$ TeV (1.1 TeV), which implies $M_{Z'} \geq 1.3$ TeV (870 GeV) and  $M_{Z''} \geq 790$ GeV (520 GeV).

The effective weak charge in atomic parity violation $Q_W$ can be used to measure the shift in the $Z$ coupling through $Z-Z'$ mixing:
\beq
Q_W = -2\left[(2Z+N)C_{1u} + (Z+2N)C_{1d}\right],
\eeq
where $C_{1q}=2g^e_Ag^q_V$. Using the deviation of the SM prediction from the experimental data for the cesium atom  $\Delta Q_W(Cs) = Q_W^{exp}-Q_W^{\textrm{SM}} = 0.55 \pm 0.49$ \cite{PDB06}, we obtain the bound 0.82 TeV $\leq f \leq$ 3.4 TeV, which also implies 0.63 TeV $\leq M_{Z^\prime} \leq$ 2.7 TeV and 0.38 TeV $\leq M_{Z^{\prime\prime}} \leq$ 1.6 TeV.

Direct experimental constraint on other four-fermion operators can also give bounds to the model. For instance, the exchange of the $Z''$ produces an operator of the size:
\beq
\eta_{LL}^{eq}\left(\bar{e}_L\gamma_\mu e_L\right)\left(\bar{q}_L\gamma_\mu q_L\right), \quad \eta_{LL}^{eq} \sim v^2/f^2.
\eeq
for left-left currents.  Using the current bounds on this operator $\eta_{LL}^{eq} = 0.01 \pm 0.20$ \cite{Cheung01}, we obtain $M_{Z^{\prime\prime}} \geq 500$ GeV, which also gives $f \geq 1.1$ TeV.

In summary, we discussed the aspects of the LHM based on the $SU(4)_L \times U(1)_X$ electroweak gauge group as an alternative solution to the naturalness and fine-tuning issues.  We introduced the anomaly-free fermion embedding in this model, and presented their interactions to the physical gauge bosons.  The new charged (flavor-changing) gauge bosons do not mix with the SM gauge bosons so that this model does not receive strong electroweak constraints in the charged sector.  On the other hand, there are two new neutral (flavor- conserving) gauge bosons $Z'$ and $Z''$ which could appear at the electroweak scale through mixing and/or directly.  Using the recent experimental values of the electroweak precision constraints, we obtained the bounds on the NP scale parameter $f$ and the masses of $Z'$ and $Z''$.

The masses of the fermions are obtained from Yukawa matrices which lead mass mixing between the ordinary and exotic fermions.  A simple way of achieving the mass splitting between them is to introduce an anomaly-free discrete symmetry.  Once the Yukawa terms are identified, one can generate the Higgs potential radiatively via the Coleman-Weinberg mechanism \cite{Coleman73}. The further detailed studies on the Higgs potential as well as the collider signatures of the NP particles are in progress \cite{prog}.


\begin{theacknowledgments}
I thank Otto Kong for his proposal and collaboration at the beginning of this project.  This work was supported in part by the National Science Council of
R.O.C. under Grant \#s:NSC-95-2112-M-006-013-MY2.
\end{theacknowledgments}

\bibliographystyle{aipproc}   

\end{document}